\documentstyle[aps,preprint]{revtex}
\begin{document}

\preprint{\vbox{\hbox{JHU-TIPAC-97015}\hbox{hep-ph/9707507}}}
 
\title{Phenomenology of New Baryons with Charm and Strangeness}
 
\author{George Chiladze and Adam F.~Falk}

\address{Department of Physics and Astronomy, The Johns Hopkins
University\\ 3400 North
Charles Street, Baltimore, Maryland 21218 U.S.A.\\ 
{\tt chila@pha.jhu.edu, falk@jhu.edu}}

\date{July 1997}

\maketitle 

\begin{abstract}%
The CLEO Collaboration recently has announced the discovery of an excited charmed and strange baryon.  We estimate the expected width of this new member of the $\Xi_c$ family.  We discuss the phenomenology of the excited $csq$ and $css$ states, and consider what additional charmed baryons might be observable in the future.  We point out that the final state $\Xi_cK$ could be an interesting new channel to examine.
\end{abstract}
\pacs{12.39.Hg, 12.40.Yx, 13.30.-a, 14.20.Lq}

\newpage 

The CLEO Collaboration recently has announced the discovery of a new excited charmed and strange baryon~\cite{CLEO1}.  The new state decays in the channel $\Xi_c^{*0}\pi^+$, which then goes to $\Xi_c^+\pi^+\pi^-$.  The measured mass difference is $M(\Xi_c^+\pi^+\pi^-)-M(\Xi_c^+)=349.4\pm0.7\pm1.0\,$MeV.  The new baryon has been identified tentatively as the strange analogue of the $\Lambda_c^+(2625)$, a baryon with total spin-parity $J^P={3\over2}^-$, because of the decay channel in which it appears.  In this letter we discuss the phenomenology of the excited charmed and strange baryons, and estimate the expected width of such a state.   We will also explore whether, and where, it would be fruitful to look for additional excited charmed baryons.

The heavy quark limit, in which $m_c,m_b\to\infty$, has proven an excellent guide to charmed and bottom hadron spectroscopy and strong decays~\cite{IW,ILW,FM,Falk1,Savage,Lebed,Jenkins}.  In this limit, the spin and parity quantum numbers of the light degrees of freedom in the hadron are conserved, and it is convenient to use them to enumerate the spectrum of states.  In addition, strong decays of excited heavy baryons are transitions solely of the light degrees of freedom, so selection rules for these decays are easily derived in terms of the light quantum numbers.  The spin of the heavy quark is then added to derive rules for physical states.   For light degrees of freedom with spin $J_\ell>0$, there exists a doublet of heavy hadrons of total spin $J=J_\ell\pm{1\over2}$.  The two states in each doublet are nearly degenerate, their masses split only by a chromomagnetic interaction, which scales as $\Lambda_{\rm QCD}^2/m_c$.  Heavy quark symmetry then relates the decay properties of the two members of each doublet.

The lowest lying charmed baryons are listed in Table~\ref{baryontable}, along with their observed masses and postulated quantum numbers.  The separate spin $s_\ell$ and orbital angular momentum $L_\ell$ of the light degrees of freedom are not well defined, and we include them simply for guidance from the quark model.  In the last column is the dominant decay mode of the state with those quantum numbers, if kinematically allowed.  The emission of $\pi$'s by excited baryons is constrained by both the $J^P$ and the $J_\ell^P$ of the initial and final state.  For example, the decay $\Lambda_{c1}(\case12)^+\to\Lambda_c^+\pi^0$ is allowed by $J^P$ but forbidden\footnote{This particular transition also happens to violate isospin.} by $J^P_\ell$; while $\Lambda_{c1}(\case12)^+\to\Sigma^{*+}_c\pi^0$ is allowed by $J^P_\ell$ but forbidden by $J^P$.  Heavy quark $SU(2)$ and flavor $SU(3)$ symmetries may be used to relate processes involving different members of a heavy quark doublet, or involving states with the same $J^P$ but different light quark content.  With these symmetries imposed, the states fill out representations of $SU(2)\times SU(3)$; in Table~\ref{baryontable} we have three such representations, labeled by $J_\ell^P$, with $J_\ell^P=0^+,1^+$ and $1^-$.  Note that Fermi statistics restricts the lightest doubly strange $\Omega_c$ states to $J_\ell^P=1^+$.

The heavy quark and light flavor symmetries relate the strong decays of the $\Lambda_{c1}$ and $\Xi_{c1}$ states.  Cho has developed a formalism, based on Heavy Hadron Chiral Perturbation Theory, for making these relations explicit, and we refer the reader to his paper for details~\cite{Cho}.  (See also the analysis of Pirjol and Yan~\cite{PY}.)  The result is that there is a single interaction in the Lagrangian which couples the $J_\ell^P=1^-$ doublet to the $J^P=1^+$ doublet and a $\pi$ in an $S$-wave.  (Decay via $D$-wave $\pi$ emission is also allowed, but it requires an operator of higher dimension which is suppressed by $|\vec p_\pi|/4\pi f_\pi$.)  The $J_\ell^P=1^-$ doublet transforms as an antitriplet under $SU(3)$, and is represented as a $J^P=\case12^-$ field $R_i$ and a $J^P=\case32^-$ Rarita-Schwinger field $R^\mu_i$.  The $J_\ell^P=1^+$ doublet transforms as a sextet under $SU(3)$, and is represented as a $J^P=\case12^+$ field $S^{kl}$ and a $J^P=\case32^+$ field $S^{kl}_\mu$.  The octet of pions and kaons appears nonlinearly in the usual axial combination $A_\mu=-\partial_\mu\pi/(\sqrt{2}f_\pi)+\cdots$, where $f_\pi\approx93\,{\rm MeV}$ is the $\pi$ decay constant.  The interaction term is
\begin{equation}
\label{term1}
  h_2\,\epsilon^{ijk}\,(v\cdot A^l_j)\,\left[\overline S_{kl} R_i
  +\overline S^{\,\mu}_{kl} R_{i\mu}\right]+\text{h.c.}\,,
\end{equation}
where $v^\mu$ is the four velocity of the heavy baryons and $h_2$ is an unknown coupling constant.  This interaction is responsible for the decay $\Lambda_{c1}(\case12)^+\to\Sigma\pi$, whose width has been measured.  (Note that the other possible decay, $\Lambda_{c1}(\case32)^+\to\Sigma^*\pi$, is not kinematically allowed.)  Since the strong decay of the $\Lambda_{c1}(\case12)^+$ occurs very close to threshold, it is important to treat the phase space exactly in each channel.

The partial decay width in a given channel due to the interaction~(\ref{term1}) is given by
\begin{equation}
\label{width}
  \Gamma(R\to S\pi)=C\,{h_2^2\over4\pi f_\pi^2}
  {M_S\over M_R}\,E_\pi^2\,|\vec p_\pi|\,,
\end{equation}
where $C$ is a group theory factor which depends on the flavor of the hadrons.  For the decay $\Lambda_{c1}(\case12)^+\to\Sigma\pi$, using the masses in Table~\ref{baryontable}, we find the partial widths\footnote{These expressions differ from those of Ref.~\cite{Cho}, due to our exact treatment of phase space.}
\begin{eqnarray}
  &&\Gamma(\Lambda_{c1}(\case12)^+\to\Sigma_c^0\pi^+)=3.9^{+1.6}_{-2.6}\,{\rm MeV}\times h_2^2\,,
  \nonumber\\
  &&\Gamma(\Lambda_{c1}(\case12)^+\to\Sigma_c^+\pi^0)=6.2^{+1.3}_{-1.4}\,{\rm MeV}\times h_2^2\,,
  \nonumber\\
  &&\Gamma(\Lambda_{c1}(\case12)^+\to\Sigma_c^{++}\pi^-)=3.0^{+1.7}_{-3.0}\,{\rm MeV}\times h_2^2\,,
\end{eqnarray}
where the errors are estimated from the uncertainty in the masses of the states.  For each of these decays, $C=1$.  Assuming that the width of the $\Lambda_{c1}(\case12)^+$ is saturated by these channels, and using the measured value $\Gamma(\Lambda_{c1}(\case12)^+)=3.9^{+2.4}_{-1.6}\,{\rm MeV}$~\cite{PDG}, we extract
$h_2^2=0.30^{+0.21}_{-0.14}$.  Given the large errors in the measured width, the neglect of other decay channels is not likely to be important in the determination of $h_2^2$.

We may then use this result and $SU(3)$ flavor symmetry to predict the width of the $\Xi_{c1}(\case32)^+$.  This state can decay in either of the channels $\Xi_c^{*0}\pi^+$ and $\Xi_c^{*+}\pi^0$.  We compute the partial widths from Eq.~(\ref{width}); for $\Xi_{c1}(\case32)^+\to\Xi_c^{*0}\pi^+$, the group theory factor is $C=\case12$, while for $\Xi_{c1}(\case32)^+\to\Xi_c^{*+}\pi^0$, $C=\case14$.  Hence we find
\begin{eqnarray}
  &&\Gamma(\Xi_{c1}(\case32)^+\to\Xi_c^{*0}\pi^+)=11.9\,{\rm MeV}\times h_2^2\,,\nonumber\\
  &&\Gamma(\Xi_{c1}(\case32)^+\to\Xi_c^{*+}\pi^0)=6.2\,{\rm MeV}\times h_2^2\,.
\end{eqnarray}
The branching ratios to $\Xi_c^{*0}\pi^+$ and $\Xi_c^{*+}\pi^0$ are predicted to be $65\%$ and $35\%$, respectively.  Assuming that these two channels saturate the decay rate, the width is then
\begin{equation}
\label{widthnum}
  \Gamma(\Xi_{c1}(\case32)^+)=5.4^{+3.8}_{-2.5}\,{\rm MeV}\,.
\end{equation}
The uncertainty in this result is dominated by the experimental error in  
$\Gamma(\Lambda_{c1}(\case12)^+)$.  Corrections due to $SU(3)$ breaking are unlikely to be larger than these already substantial uncertainties.  We see that the $\Xi_{c1}(\case32)$ is expected to be quite narrow, despite decaying via $S$-wave $\pi$ emission.  The predicted width (\ref{widthnum}) is reasonably consistent with the upper bound obtained by CLEO, $\Gamma(\Xi_{c1}(\case32)^+)<2.4\,$MeV at the 90\% confidence level.  In particular, this result supports the identification of the new state as the $\Xi_{c1}(\case32)$.

By now, most of the low lying charmed baryons listed in Table~\ref{baryontable} have been discovered.  Only a few more states are needed to complete the task.  The isospin partner of the $\Xi_{c1}(\case32)^+$ should not prove much more difficult to find, since it has a decay into charged pions, 
$\Xi_{c1}(\case32)^0\to\Xi_c^{*+}\pi^-\to\Xi_c^0\pi^+\pi^-$.  However, the other member of the heavy doublet, the $\Xi_{c1}(\case12)$, decays to $\Xi_c'$, which itself decays radiatively and is hard to isolate.  Similarly, the $\Omega_c^*$ decays radiatively to $\Omega_c$ and also may be difficult to find.  Finally, there is the $\Sigma_c^{*+}$, which, unlike its isospin partners $\Sigma_c^{*0}$ and $\Sigma_c^{*++}$, decays to a neutral pion and has not yet been identified. 

What other excited charmed baryons might one look for?  The lightest undiscovered states are likely to be of two types.  First, there are  ``radial'' excitations of the ground state $\Lambda_c$ and $\Xi_c$, which have $J_\ell^P=0^+$ and $J^P=\case12^+$.  Second, there are ``orbital'' excitations of the $\Sigma_c^{(*)}$, $\Xi_c'^{(*)}$ and $\Omega_c^{(*)}$, which have $s_\ell=L_\ell=1$ (in the quark model), and hence $J_\ell^P=0^-,1^-$ or $2^-$.  If the sign of the spin-orbit interaction follows one's quark model intuition, then the lightest of these ``orbital'' excitations will have $J_\ell^P=0^-$ and $J^P=\case12^-$.

It is hard to estimate the widths of these excited states with any precision, both because their strong decays proceed via nonperturbative interactions and because their masses, and hence the available phase space, are not known.  The best that we can do is to note that the decays will be mediated by operators analogous to Eq.~(\ref{term1}), with new coupling constants of order one replacing $h_2$.  We can then expand the width for $E_\pi\approx m_\pi$, since to be visible a new state would have to be fairly close to threshold for decay via $\pi$ emission.  For $S$-wave decays, we can scale directly from Eq.~(\ref{width}), while for $P$-wave decays we make the replacement $E_\pi^2|\vec p_\pi|\to|\vec p_\pi|^3$, and for $D$-wave decays $E_\pi^2|\vec p_\pi|\to|\vec p_\pi|^5/(4\pi f_\pi)^2$.  Neglecting constants of order one, we find that near threshold,
\begin{mathletters}
\label{pionwidths}
\begin{eqnarray}
  \label{swavepi}
  S{\rm -wave}:&\qquad&\Gamma\sim10\,{\rm MeV}\times
  \left[{E_\pi-m_\pi\over100\,{\rm MeV}}\right]^{1/2}\,,\\
  \label{pwavepi}
  P{\rm -wave}:&\qquad&\Gamma\sim10\,{\rm MeV}\times
  \left[{E_\pi-m_\pi\over100\,{\rm MeV}}\right]^{3/2}\,,\\
  \label{dwavepi}
  D{\rm -wave}:&\qquad&\Gamma\sim1\,{\rm MeV}\times
  \left[{E_\pi-m_\pi\over100\,{\rm MeV}}\right]^{5/2}\,.
\end{eqnarray}
\end{mathletters}
These expressions should be taken only as rough guesses intended to provide guidance as to the mass range which would make a new state narrow enough to be observable.

Let us denote the ``radial'' excitations by $\Lambda_c^{**}$ and $\Xi_c^{**}$.  If their mass is large enough, they will decay in the channels $\Lambda_c^{**}\to\Sigma_c\pi$ and $\Xi_c^{**}\to\Xi'_c\pi$, where the $\pi$ is emitted in a $P$-wave.  According to Eq.~(\ref{pwavepi}), these states will be broad, unless they are fairly close to threshold for $\pi$ emission.  If they are lighter than this, then the $\Lambda_c^{**}$ and $\Xi_c^{**}$ will decay radiatively or via $\pi\pi$ to $\Lambda_c$ and $\Xi_c$, in which case they will be narrow.  Since there is no firm theoretical prediction for the masses of the $\Lambda_c^{**}$ and $\Xi_c^{**}$, it is worth searching for them in these channels, in the hope that they will be fairly light.  However, the $\Sigma_c\pi$ channel is already well studied in this mass range, which may bode ill for future discoveries, and states which decay to $\Xi_c'$ will be hard to find.  Although the decay $\Xi_c^{**}\to\Xi_c^*\pi$ may also be possible, it will not compete with $\Xi_c^{**}\to\Xi_c'\pi$ unless both partial widths are large, and thus the $\Xi_c^{**}$ is broad.

The ``orbital'' excitations are potentially more interesting.\footnote{The properties of these excitations are also discussed in Ref.~\cite{PY}.}  Let us denote the heavy states by $\Sigma_{ci}$, $\Xi_{ci}^*$ and $\Omega_{ci}$, where $i=0,1,2$ for $J_\ell^P=0^-,1^-,2^-$.  For $J_\ell^P\ne0$ the physical states are a heavy doublet of spin $J=J_\ell^P\pm\case12$.  Although one might expect, based on the quark model, that the masses of these $s_\ell=L_\ell=1$ states would not be smaller than those of the observed $s_\ell=0$ and $L_\ell=1$ baryons, the unknown strength of the spin-orbit interaction makes them difficult to estimate.

The $\Sigma_{c0}$ baryon can decay in the channel $\Lambda_c\pi$, where the $\pi$ is emitted in an $S$-wave.  From Eq.~(\ref{swavepi}) we see that unless it is very light, this state is likely to be too broad to be observable.  The $\Sigma_{c1}$ doublet has the same quantum numbers as the observed $\Lambda_{c1}$ states, and will decay according to the same pattern, namely to $\Lambda_c\pi\pi$, resonating through $\Sigma_c^{(*)}$ if kinematically allowed.  Finally, the $\Sigma_{c2}$ doublet can decay to $\Lambda_c\pi$ through emission of a $D$-wave $\pi$. The $\Xi^*_{ci}$ baryons will be similar to the $\Sigma_{ci}$.  The $\Xi^*_{c0}$ is probably broad,  the $\Xi_{c1}^*$ doublet is analogous to the $\Xi_{c1}$, and the $\Xi^*_{c2}$ decays to $\Xi_c\pi$ in a $D$-wave.  Note that in each flavor sector, corrections proportional to $1/m_c$ can mix the two $J^P=\case32^-$ baryons with $J^P_\ell=1^-$ and $2^-$, and the two $J^P=\case12^-$ states with $J^P_\ell=0^-$ and $1^-$.

From the width estimate (\ref{dwavepi}), we see that the $|\vec p_\pi|^5$ suppression in the $D$-wave decay widths might make the $\Sigma_{c2}$ and $\Xi^*_{c2}$ states fairly narrow.  For example, in the charmed meson sector the $J^P=2^+$ $D_2(2460)$ decays to $D$ and $D^*$ via $D$-wave $\pi$ emission, and although the mass differences are $590\,$MeV and $450\,$MeV, respectively, the width of the $D_2(2460)$ is only approximately $20\,$MeV.  However, its $J^P=1^+$ heavy partner, the $D_1(2420)$, is broadened substantially by effects of order $1/m_c$, perhaps by  mixing with a wide state with the same $J^P$ but different $J^P_\ell$ which decays by emitting a $\pi$ in an $S$-wave~\cite{ILW,FM}.  Similarly, the $J^P=\case32^-$ members of the $\Sigma_{c2}$ and $\Xi_{c2}$ doublets could be broadened by mixing with the $J^P=\case32^-$ states of the $\Sigma_{c1}$ and $\Xi_{c1}$.  Hence the pure $D$-wave decays of the $J^P=\case52^-$ baryons might be the easiest to find.  Finally, note that the partial width of the $\Xi^*_{c2}$ into $\Xi_c\pi$ will be suppressed further by an isospin factor of $\case14$.  These are decay channels worth exploring, in the mass ranges $M(\Lambda_c\pi)\agt2420\,$MeV and $M(\Xi_c\pi)\agt2500\,$MeV.

The $\Omega_{ci}$ states decay somewhat differently, because of the absence of light quarks.  If they are too light, they can decay only radiatively to $\Omega_c^{(*)}$ and will be very hard to find.  If its mass is greater than $2960\,$MeV, the $\Omega_{c0}$ will decay to the final state $\Xi_cK$. The $\Omega_{c1}$ cannot decay to $\Xi_cK$ because of parity; instead, it will go to $\Omega_c\pi\pi$ if $M(\Omega_{c1})>2972\,$MeV, and to $\Xi_c'K$ if $M(\Omega_{c1})>3075\,$MeV.  The best hope for seeing this state is if $\Omega_{c1}\to\Omega_c\pi\pi$ is the only strong decay allowed.  The $\Omega_{c2}$ will decay to $\Xi_cK$ if possible, with the $K$ in a $D$-wave.  As above, the $J^P=\case32^-$ member of this doublet might be broadened by $1/m_c$ effects.

Since a mass in the vicinity of $M(\Xi_c)+M(K)\approx2960\,$MeV is quite possible for the $\Omega^*_{ci}$ states, the decay channel $\Xi_cK$ could be a promising place to look.  For $S$-wave decays, an estimate similar to Eq.~(\ref{swavepi}) applies, but the growth of $\Gamma$ away from threshold is much faster for a $K$ in the final state than for a $\pi$, because near threshold $E_K^2\approx m_K^2\gg m_\pi^2$.  Neglecting constants of order one, 
the rough estimates for kaons analogous to Eq.~(\ref{pionwidths}) are
\begin{mathletters}
\begin{eqnarray}
  \label{swavek}
  S{\rm -wave}:&\qquad&\Gamma\sim100\,{\rm MeV}\times
  \left[{E_K-m_K\over10\,{\rm MeV}}\right]^{1/2}\,,\\
  \label{pwavek}
  P{\rm -wave}:&\qquad&\Gamma\sim10\,{\rm MeV}\times
  \left[{E_K-m_K\over10\,{\rm MeV}}\right]^{3/2}\,,\\
  \label{dwavek}
  D{\rm -wave}:&\qquad&\Gamma\sim10\,{\rm MeV}\times
  \left[{E_K-m_K\over100\,{\rm MeV}}\right]^{5/2}\,.
\end{eqnarray}
\end{mathletters}
Thus the $\Omega^*_{c0}$, which decays in an $S$-wave, will likely be too wide to be observable unless it lies within $1\,$MeV or so of threshold.  By contrast, the $\Omega_{c2}^*$ is likely to be reasonably narrow even if it is as heavy as $3100\,$MeV.  Assuming that enough of them can be produced, this state should be a prime candidate for discovery in the $\Xi_cK$ channel.

We are grateful to John Yelton for discussions.  This work was supported in part by the National Science  Foundation under Grant No.~PHY-9404057.  A.F.F.~was also supported by the National Science  Foundation under National Young Investigator Award No.~PHY-9457916, by the Department of Energy under Outstanding Junior Investigator Award No.~DE-FG02-94ER40869, and by the Alfred P.~Sloan Foundation.  A.F.F. is a Cottrell Scholar of the Research Corporation.

\begin{table}
  \begin{tabular}{lcccccccc}
  Name&Mass (MeV)&$J^P$&$J_\ell^P$&``$s_\ell$''&``$L_\ell$''&$I$&$S$&Dominant Decay\\
  \hline
  $\Lambda_c^+$&$2284.9\pm0.6$&$\case12^+$&$0^+$&0&0&0&0&weak\\
  $\Sigma_c^0$&$2452.1\pm0.7$&$\case12^+$&$1^+$&1&0&1&0&$\Lambda_c^+\pi^-\,(P)$\\
  $\Sigma_c^+$&$2453.5\pm0.9$&$\case12^+$&$1^+$&1&0&1&0&$\Lambda_c^+\pi^0\,(P)$\\
  $\Sigma_c^{++}$&$2452.9\pm0.6$&$\case12^+$&$1^+$&1&0&1&0&$\Lambda_c^+\pi^+\,(P)$\\
  $\Sigma^{*0}_c$&$2517.5\pm1.6$&$\case32^+$&$1^+$&1&0&1&0&$\Lambda_c^+\pi^-\,(P)$\\
  $\Sigma^{*+}_c$&?&$\case32^+$&$1^+$&1&0&1&0&$\Lambda_c^+\pi^0\,(P)$\\
  $\Sigma^{*++}_c$&$2519.4\pm1.6$&$\case32^+$&$1^+$&1&0&1&0&$\Lambda_c^+\pi^+\,(P)$\\
  $\Lambda_{c1}(\case12)^+$&$2593.6\pm1.0$&$\case12^-$&$1^-$&0&1&0&0&$\Sigma_c\pi\,(S)$\\
  $\Lambda_{c1}(\case32)^+$&$2626.4\pm0.9$&$\case32^-$&$1^-$&0&1&0&0&$\Lambda_c^+\pi\pi\,(S,P)$\\
  $\Xi_c^0$&$2470.3\pm1.8$&$\case12^+$&$0^+$&0&0&$\case12$&$-1$&weak\\
  $\Xi_c^+$&$2465.6\pm1.4$&$\case12^+$&$0^+$&0&0&$\case12$&$-1$&weak\\
  $\Xi'^{\,0,+}_c$&$\sim2580$ (?)&$\case12^+$&$1^+$&1&0&$\case12$&$-1$&$\Xi_c\gamma$\\
  $\Xi^{*0}_c$&$2643.8\pm1.8$&$\case32^+$&$1^+$&1&0&$\case12$&$-1$&$\Xi_c\pi\,(P)$\\
  $\Xi^{*+}_c$&$2644.6\pm2.1$&$\case32^+$&$1^+$&1&0&$\case12$&$-1$&$\Xi_c\pi\,(P)$\\
  $\Xi_{c1}(\case12)^{\,0,+}$&?&$\case12^-$&$1^-$&0&1&$\case12$&$-1$&$\Xi_c'\pi\,(S)$\\
  $\Xi_{c1}(\case32)^0$&?&$\case32^-$&$1^-$&0&1&$\case12$&$-1$&$\Xi_c^*\pi\,(S)$\\
  $\Xi_{c1}(\case32)^+$&$2815.0\pm1.9$&$\case32^-$&$1^-$&0&1&$\case12$&$-1$&$\Xi_c^*\pi\,(S)$\\
  $\Omega_c^0$&$2704\pm4$&$\case12^+$&$1^+$&1&0&0&$-2$&weak\\
  $\Omega_c^{*0}$&?&$\case32^+$&$1^+$&1&0&0&$-2$&$\Omega_c\gamma$\\
  \end{tabular}
\vspace{0.4cm}
  \caption{The lowest lying charmed baryons~\protect\cite{PDG,CLEO2,CLEO3,CLEO4}.  The spin and orbital angular momentum quantum numbers $s_\ell$ and $L_\ell$ are only defined in the quark model, and are included here for guidance.  Isospin and strangeness are denoted respectively by $I$ and $S$.  The angular momentum of the $\pi$ emitted in a strong decay is indicated.  For simplicity, we give estimated errors on baryon masses themselves, rather than reporting the (more accurately measured) mass differences.}
  \label{baryontable}
\end{table}

\end{document}